# Direct electrochemical reduction of graphene oxide thin film for aptamer-based selective and highly sensitive detection of Matrix metalloproteinase 2

Stefan Jarić[a*], Silvia Schobesberger[b], Lazar Velicki[c,d], Aleksandra Milovančev[c,d], Stanislava Nikolić[c,e] Peter Ertl[b], Ivan Bobrinetskiy[a], and Nikola Ž. Knežević[a*]

[a]Biosense Institute – Research and Development Institute for Information Technologies in Biosystems, University of Novi Sad, Dr Zorana Đinđića 1, 21000 Novi Sad, Serbia

[b]TU Wien, Faculty of Technical Chemistry, Getreidemarkt 9, 1060 Vienna, Austria

[c]Faculty of Medicine, University of Novi Sad, Hajduk Veljkova 3, 21000, Novi Sad, Serbia

[d]Institute of Cardiovascular Diseases of Vojvodina, Put Doktora Goldmana 4, 21204, Sremska Kamenica, Serbia

[e]Center of Laboratory Medicine, Clinical Center of Vojvodina, Hajduk Veljkova 1, 21000, Novi Sad, Serbia

*Corresponding authors; e-mails: sjaric@biosense.rs, nknezevic@biosense.rs

## Abstract

Simple and low-cost biosensing solutions are suitable for point-of-care applications aiming to overcome the gap between scientific concepts and technological production. To compete with sensitivity and selectivity of golden standards, such as liquid chromatography, the functionalization of biosensors is continuously optimized to enhance the signal and improve their performance, often leading to complex chemical assay development. In this research, the efforts are made on optimizing the methodology for electrochemical reduction of graphene oxide to produce thin film-modified gold electrodes. Under the employed specific conditions, 20 cycles of cyclic voltammetry (CV) are shown to be optimal for superior electrical activation of graphene oxide into electrochemically reduced graphene oxide (ERGO). This platform is further used to develop a matrix metalloproteinase 2 (MMP-2) biosensor, where specific anti-MMP2 aptamers are utilized as a biorecognition element. MMP-2 is a protein which is typically overexpressed in tumor tissues, with important roles in tumor invasion, metastasis as well as in tumor angiogenesis. Based on impedimetric measurements, we were able to detect as low as 3.32 pg mL$^{-1}$ of MMP-2 in PBS with a dynamic range of 10 pg mL$^{-1}$ – 10 ng mL$^{-1}$. Further experiments with real blood samples revealed a promising potential of the developed sensor for direct measurement of MMP-2 in complex media. High specificity of detection is demonstrated – even to the closely related enzyme MMP-9. Finally, the potential of reuse was demonstrated by signal restoration after experimental detection of MMP-2.

## Graphical abstract

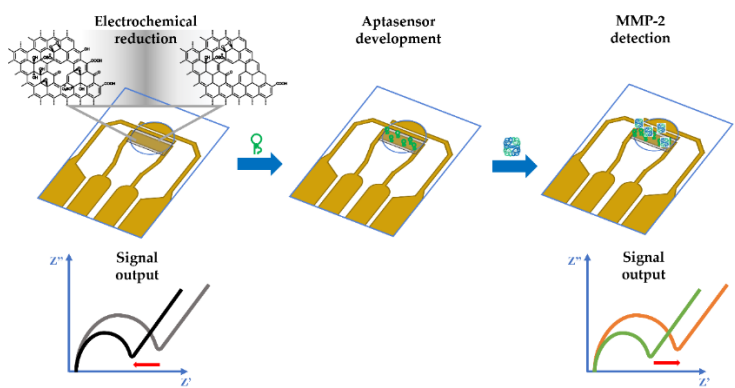

**Introduction**

Matrix metalloproteinases (MMPs) are enzymes of a broad spectrum of physiological and pathological processes primarily in an extracellular matrix (ECM) [1]. Increased MMP activities are known to occur in localized abnormal physiological conditions, such as cancer, enhancing angiogenesis but having also pro- or anti-angiogenic effect in cancer tissue and making them angiomodulators [2]. Matrix metalloproteinase 2 (MMP-2) or 72 kDa type-IV gelatinase is a type of MMP known for the degradation of collagen. MMP-2 is significantly involved in many pathological processes, especially in carcinogenic-related diseases, for example, in prostate carcinogenesis [3] or breast carcinogenesis [4]. Because MMP-2 is evidenced to be overexpressed due to many different diseases, it is not considered as a specific biomarker of a disease, but rather as a prognostic biomarker [5].

The critical point is its concentration found in human fluids from healthy and diseased individuals. Morgia et al. evaluated MMP-2 activity (activated MMP-2) in patients diagnosed with prostate cancer (PC) from plasma samples [5]. Namely, a control group of people had a mean concentration of 14.3 ng mL$^{-1}$, patients with benign prostate hyperplasia (BPH) had 15.4 ng mL$^{-1}$, while patients with PC had 24.2 ng mL$^{-1}$ (no metastasis) and 82.4 ng mL$^{-1}$ (with metastasis). For a comparison, they additionally quantified total MMP-2 concentrations in plasma and determined 471.7 ng mL$^{-1}$, 814.0 ng mL$^{-1}$, 893.0 ng mL$^{-1}$, and 1230.0 ng mL$^{-1}$ for control, BPH, non-metastatic PC, and metastatic PC, respectively. This means that only 2-7% of total MMP-2 is active in plasma, with the highest activity found in patients with metastatic PC. Moreover, the noninvasive methods have to be developed for PoC prognostic MMP2 analysis, e.g., in urine where concentration can be less than 1 ng mL$^{-1}$ [6]. This implies that enzymatically active MMP-2 is present in lower concentrations (<1 ng mL$^{-1}$), which reveals the need to develop ultra-sensitive biosensors for MMP-2.

Electrochemical (EC)-based biosensors have been shown to be very sensitive for protein detection, especially when novel technologies based on nanomaterials (magnetic beads, gold nanoparticles (AuNPs), graphene, quantum dots, and carbon nanotubes) are employed [7]. The ultrasensitivity of graphene-based biosensors for biomarkers (protein) detection is reported by many research groups [8,9,10] due to unprecedented electrical properties of graphene as well as high surface-to-volume ratio. One of the EC solutions for MMP-2 detection relies on the use of graphene-oxide micro-flakes functionalized with horseradish peroxidase (HRP) [11]. The primary antibodies for MMP-2 were immobilized onto AuNP/N-doped graphene composites, while the secondary antibodies were immobilized onto polydopamine-modified GO, enabling detection of 0.11 pg mL$^{-1}$ MMP-2. In another study, aptamer-based biosensor for the detection of MMP-2 is developed based on EC methodology utilizing novel aptamer immobilization strategy [12]. Namely, IDE gold electrodes were functionalized with monolayer graphene, which was modified with anti-MMP-2 aptamers. By using differential pulse voltammetry (DPV), a limit of detection was determined to be 100 pM (6.2 ng mL$^{-1}$). Development of novel methods for improved biosensing, that are based on the conjugation of nanomaterials and multiple chemical units enabling more sensitive conversion of chemical interaction into physical signal, can meet challenges when it comes to the technology scale-up. Therefore, low-cost, and simple assay biosensing solutions, yet reliable, are emerging in this field to meet the future possibilities of commercialization. The development of aptamer-functionalized graphene-based sensors is a particularly promising approach for production of efficient biosensors for health-relevant biomarkers [13].

Graphene and its derivatives are extensively used 2D nanomaterials in biosensing technologies [14,15]. Pristine monolayer graphene or graphene oxide (GO) is weakly selective when used in biosensors but owing to their possibility for surface modifications and significant improvement in biosensing specificity, they are prominent in biosensor development [16]. The challenging side of using graphene materials is of technological nature, however, the integration of these materials can be improved to preserve their extraordinary properties. Technically more practical GO is a good substitute for monolayer graphene because it contains oxide groups and defects, which ensure hydrophilicity, chemical and electrochemical activity. We suggest using GO as a sensitive layer due to robust, cost-effective, and scalable technology for biosensors manufacture.

GO is a single carbon layer decorated with oxygen groups on the basal plane and its edges. It often contains defects, which gives GO hybrid structure composed of $sp^2$- and $sp^3$-hybridized carbon atoms resulting in different properties such as fluorescence quenching, biocompatibility, facile chemical modification but low electrical conductivity [17]. Low conductivity of GO can be impractical for electrochemical biosensing and there are different strategies for GO reduction, such as chemical, electrochemical, or thermal reduction. EC reduction of GO is a simple and environmentally friendly technique that uses a certain negative potential to reduce/remove oxide groups. The negative potential can be applied as a constant over reduction time or by a voltammetric technique [18], in most cases CV, to obtain electrochemically reduced GO (ERGO). The advantage of EC reduction includes its operation at room temperature and in buffer electrolyte but also that it can be performed from colloidal suspension of GO or in a three-electrode system where working electrode is modified with a GO thin film. It is important to note that CV gives real-time information about redox potentials and the amount of reduction during the process [19].

Herein, the EC reduction protocol of GO towards yielding an ERGO thin film on commercial gold electrodes is optimized. Further surface functionalization with suitable aptamers is achieved and demonstrated for selective and highly sensitive impedimetric detection of MMP-2. A CV reduction protocol was optimized to improve electrical performance of GO, enabling the development of a sensitive and selective electrochemical biosensor for MMP-2 detection.

**Experimental section**

*Materials & Equipment*

Graphene oxide (2 mg ml$^{-1}$, dispersion in $H_2O$); potassium chloride (≥99.0%); potassium ferricyanide (≥99.0%); magnesium chloride hexahydrate (99.0-102.0%); Dulbecco's phosphate buffered saline, modified without $CaCl_2$ and $MgCl_2$; 1-pyrenebutanoic acid N-hydroxysuccinimide ester (PBASE) (95%); N,N-dimethylformamide, anhydrous (DMF) (99.8%); 1-methyl 2-pyrrolidinone (NMP) (≥99.0%); (3-aminopropyl)triethoxysilane (APTES) (≥99.0%); β-mecaptoethylamine (MEA, cysteamine) (≥98.0% RT) were all purchased from Sigma-Aldrich (USA). Matrix metalloproteinase-2 (MMP-2) and matrix metalloproteinase 9 (MMP-9), human recombinant, pre-activated, were purchased from Sigma-Aldrich (USA). Ethanol, absolute was obtained from Chem-Lab NV (Belgium). Potassium ferrocyanide trihydrate (≥99.5%) was purchased from Fluka (Spain). MMP2-specific aptamer of the sequence [20]: 5'-$NH_2(CH_3)_2$-TTT TTT TCG CCG TGT AGG ATT AGG CCA GGT ATG GGA ACC CGG TAA C-3' was synthesized at Metabion

International AG (Germany). Human Her2/ErbB2 protein, His Tag (MALS verified) (>95.0% SDS-PAGE) was purchased from ACRO Biosystems (USA). Fibrinogen (FBG) was purchased from CSL Behring (USA) and STA® - QUALI-CLOT I kit was purchased from Stago (France). Ultrapure water was produced by Barnstead™ Smart2Pure™ water purification system, 12 UV/UF (ThermoFisher Scientific, USA). Commercial gold interdigitated electrodes (IDE) ED-IDA1-Au of 10/10 μm width/gap dimension were bought from Micrux Technologies (Spain). Human MMP-2 ELISA Kit (Cat. No. KHC3081, Bender MedSystems GmbH) was purchased from Thermo Fisher Scientific (USA). Serum samples were obtained from Institute of Cardiovascular Diseases of Vojvodina (ICVDV), Novi Sad, Serbia.

All electrochemical measurements (cyclic voltammetry (CV), electrochemical impedance spectroscopy (EIS)) were performed on Biologic VMP-3e Multichannel Potentiostat (France) and EC-lab software. The electrolyte with redox probe used for all electrochemical experiments was 0.1 M KCl with 5 mM $K_3[Fe(CN)_6]/K_4[Fe(CN)_6]$ (ferro/ferricyanide redox probe). Raman spectroscopy was performed on XploRA Plus Raman Spectrometer (Horiba, France), using the 532 nm laser source, 100x objective (Olympus, Japan) (NA 0.9), and laser power 0.9 mW. Atomic force microscopy (AFM) (Solver-PRO, NT-MDT, Russia) was used to estimate the GO thickness and surface coverage.

*Graphene oxide deposition and electrochemical reduction*

Prior to gold IDE modification, each electrode was washed in ethanol and DI water with ultrasonication for 5 minutes, and finally dried with pressurized air. Washed electrodes were pre-conditioned using CV between -0.5 and 1.0 V with a scan rate 0.5 V s$^{-1}$ in 0.05 M $H_2SO_4$. To modify the surface with positively charged amino groups, silanization of the glass surface with APTES was performed by incubation with 2% APTES in ethanol for 1 hour. After rinsing with pure ethanol, electrodes were dried and annealed at 120 °C for 1 hour. The gold surface was then modified with a 10 mM MEA aqueous solution by incubating for 1 hour. After rinsing with DI water, IDEs were dried and prepared for GO deposition. GO suspension was diluted with NMP to 0.2 mg mL$^{-1}$ and homogenized by sonication. To produce ultra-thin films of GO, 10 μL droplets were cast over working electrodes and left for 2 hours at RT, ensuring negligible solvent evaporation to produce the film as uniform as possible. After incubation, the GO droplet was removed, and electrodes were annealed in low-temperature conditions (100 °C) for 30 minutes to evaporate the excess of solvent. Thin-film GO was reduced electrochemically by running 20 cycles of CV in 1X PBS using voltage range between -1.2 and -0.4 V (0.05 V s$^{-1}$ scan rate).

*Biosensor assembly*

Graphene-based biosensor was developed using our previous functionalization protocol [21]. Briefly, ERGO-IDE working electrodes were activated with 5 mM PBASE solution in DMF for 3 hours in a dark environment to enable aptamer immobilization. After incubation, the electrode was washed with fresh DMF to remove non-reacted PBASE molecules, and thoroughly washed with 2-propanol and DI water to remove the DMF residuals. Next, the aptamer stock solution (100 μM) was diluted with DPBS with 1 mM $MgCl_2$ to an intermediate concentration for proper unfolding and later diluted to a working concentration of 0.5 μM. The aptamer solution was drop-casted over electrode and incubated overnight at 4 °C in humid atmosphere to avoid solvent evaporation. The electrode was properly washed with PBS to remove unbound aptamers afterwards. Unreacted ester groups of PBASE linker were blocked with 100 mM ethanolamine solution in 1X PBS. A blocking buffer (0.05% BSA in 1X PBS) was used to block the remaining ERGO surface and hinder unspecific reactions.

*Biosensor validation*

MMP-2 stock solution was serially diluted in 1X PBS solution to obtain working concentrations ranging from 1 pg ml$^{-1}$ to 100 ng ml$^{-1}$. To evaluate the sensitivity of the biosensing platform, EIS measurements were taken for each concentration, as well as blank measurement in pure 1X PBS. Prior to each impedance measurement in the frequency range 0.25 – 100,000 Hz (10 mV potential amplitude), 20 µL of the analyte was incubated for 15 minutes, then the electrode was rinsed with ultrapure water, electrolyte solution with 5 mM ferro/ferricyanide redox probe was introduced, and measurement was performed three times (average value was taken). Biosensor signal is calculated as S = ($R_{ct,i}$ – $R_{ct,b}$)/$R_{ct,b}$ * 100 (in %), where $R_{ct,i}$ is a fitted impedance obtained from Nyquist plot of the certain analyte concentration and $R_{ct,b}$ is a fitted impedance of the blank measurement.

For the cross-selectivity measurements, anti-MMP2 aptamer-functionalized biosensor was tested against MMP-9, FBG, and ErbB2 proteins prepared also in 1X PBS. The measurements were performed in the same manner as explained above.

Biosensor regeneration tests were performed after each series of measurement to study the possibility of the biosensor re-use. For that manner, 10 mM NaOH solution was used to thoroughly wash out MMP-2 after final analyte concentration and restore aptamer configuration [22,23]. After the washing process, 1X PBS solution was again incubated and blank EIS spectrum recorded.

Blood samples were taken from three diabetic patients and serum samples were prepared. To evaluate the concentration of MMP-2 in real samples, ELISA test was performed according to manufacturer's procedures (Table S1). Each patient sample was tested twice, and the average concentration of MMP-2 is presented. On the other hand, to validate our device performance in complex matrices, serum samples were initially diluted 10-fold with 1X PBS to obtain concentrations within dynamic range of the proposed biosensor. The obtained signal S is converted to an MMP-2 concentration by employing the calibration curve and the determined concentration compared to the values obtained from ELISA.

**Results and discussion**

*Graphene oxide electrochemical reduction optimization*

To maximize the surface coverage with positively charged amino groups and achieve full coverage of GO and a homogeneous film on the working electrodes, the self-assembled monolayers (SAMs) of MEA on Au electrodes was conducted, while functionalization of the glass substrate was achieved by silanization with APTES [24]. Since oxide groups of GO are negatively charged, their attachment to the amino-functionalized positively charged sensing platform was successfully achieved by drop-casting methodology.

The working solution of 0.2 mg mL$^{-1}$ GO monolayer sheets (<10 µm diameter) was prepared by dilution with NMP in a 1:9 ratio. The obtained GO suspension in the water/NMP mixture was demonstrated to be stable for months after short sonication [25]. Thus, prior to use, GO suspension was briefly redispersed in ultrasonic bath for 5 minutes to obtain a homogeneous solution. As the GO suspension in the employed water/NMP mixture has a pH about 7.3, and the pKa of an amino group is 10.75, consequently, below a pH of 10.75, the amino group is protonated, i.e., positively charged [26], which is expected to lead to strong GO adsorption and high stability of the assembly. The full electrode preparation is presented schematically in Figure 1.

A GO thin film was successfully deposited, which can be recognized by the CV and EIS signal change (Figure 1C and D). The CV oxidation peak is significantly reduced, when considering the current magnitude, but also shifted toward higher potentials. Similarly, Nyquist plots of the GO-functionalized assembly show highly increased $R_{ct}$ resistance compared to the bare gold electrode. Evidently, a simple method of drop-casting deposition significantly impacted the electrochemical characteristics of the electrodes: the most conductive bare Au IDE has converted to the lowest conductive after GO coating (see insets of Figure 1C and D and Figure S1).

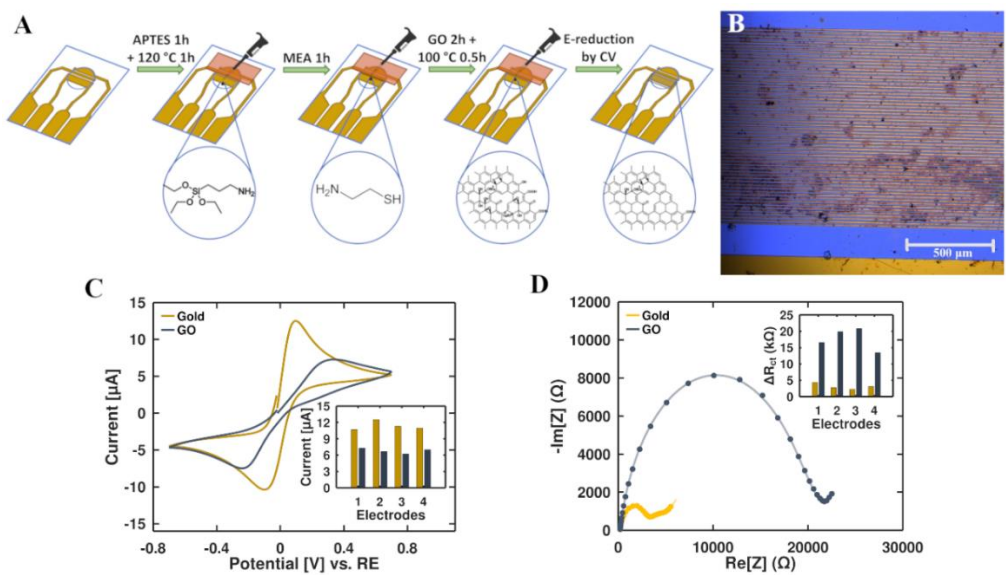

**Figure 1.** (A) Schematical illustration of the electrode modification with GO thin film and subsequent reduction; (B) Optical image of IDE working electrodes after modification; (C) CV voltammograms of a bare gold electrode and GO-coated gold electrode, and resulting oxidation peak current for four electrodes (inset); (D) Nyquist plots of a bare gold electrode and GO-coated gold electrode, and $R_{ct}$ values obtained from equivalent Randles circuit model fit for four electrodes (inset)

Although dielectric properties of as-deposited GO thin film are considered as a significant drawback, particularly for electrochemical biosensing applications, various strategies for reduction of GO have been explored for over a decade. As an example, EC reduction is a practical, easy, and low-cost method to produce conductive GO films, e.g., ERGO films without the use of aggressive and hazardous chemistry or high temperatures and inert atmosphere [24,27,28].

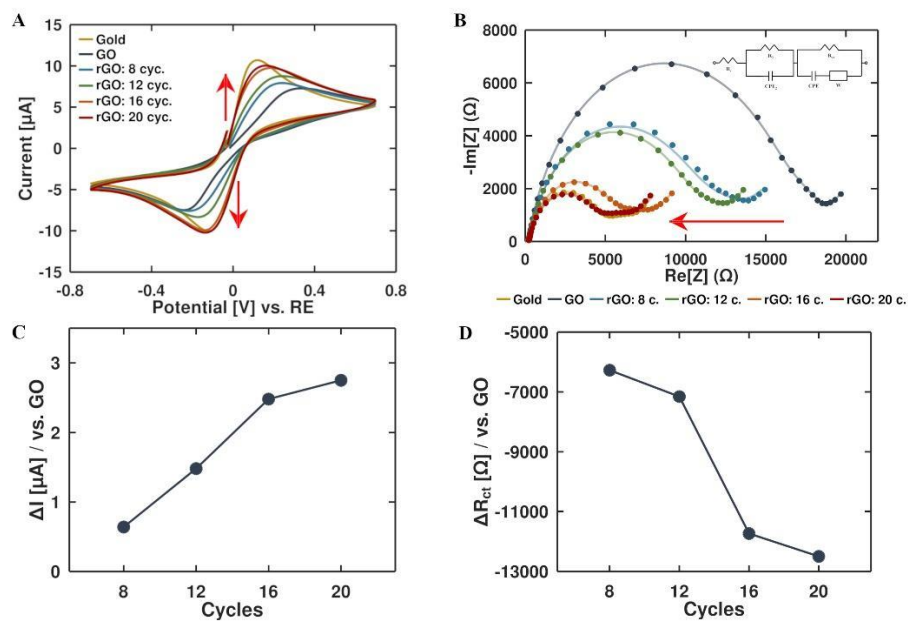

**Figure 2**. GO reduction optimization process by increasing the number of CV cycles and characterized via (A) CV, where arrows indicate peak current change, and (B) EIS, where arrow indicates impedance change, inset: Equivalent Randles circuit used to fit EIS spectra; (C) Extracted oxidation peak current change due to the CV cycles number variation referred to GO; (D) Change of $R_{ct}$ fitted value due to the CV cycles number variation referred to GO

Here, we applied CV cycling in a negative potential range (from -0.4 to -1.2 V) in 1X PBS solution with a scan rate of 50 mV s$^{-1}$ to study the reduction process of deposited GO on gold IDEs [27]. A typical voltammogram of the CV reduction protocol is characterized by a broad cathodic current peak in the first cycle present in the potential range of -0.4 to -1.2 V [16]. In our system, a first-cycle cathodic peak is observed in the potential range of -0.8 to -1.0 V, see Figure S2. This peak is diminishing with each cycle since there is no counter oxidation peak to compensate for the reduction process. Reduction of oxide groups in aqueous solutions is advantageous due to the narrowed potential windows of oxide group reduction and easier access of electrons to oxide sites [24]. We studied the number of CV cycles required to reduce GO to obtain an overall increased electrochemical performances of modified gold electrodes compared to bare gold electrodes. A set of CVs with 8, 12, 16, and 20 cycles with the same parameters were measured to compare electrochemical parameters of ERGO. In Figure 2A, CV voltammograms of ERGOs produced from different number of CV cycles are shown and compared with the bare gold and the GO-modified electrode. A low number of cycles initiate reduction, but the oxidation peak is slightly increased compared to GO. At a higher number of cycles, the oxidation peak of ERGO is significantly increased and comparable to the one of bare gold. A plot of oxidation peak current with number of cycles is extracted and slight saturation is observed at 20 cycles, Figure 2C. We assume that use of IDE configuration increases the efficiency of reduction process due to easy removal of by-products (CO, $CO_2$, and $H_2O$ [29]) between electrodes by lateral component of induced electrical field. We also applied EIS measurement to characterize ERGOs (Figure 2B) and observed a similar trend. The Nyquist semi-circle is diminishing with the increase of cycle numbers, suggesting the overall impedance reduction and same saturation of $R_{ct}$ resistance is observed at 20 cycles, as shown in Figure 2D. This leads to the conclusion that for GO-coated gold IDEs, 20 CV cycles are optimal to produce ERGO with same or better electrochemical parameters than bare gold IDEs. This is important from the perspective of the use of GO as nanomaterial of choice for biosensor development. Stability of ERGO films on the sensing surface is possibly supported by strong interaction of remaining oxide groups of GO with SAMs on gold surface, with the existence of π-π interaction between ERGO and GO.

**Characterization of GO and ERGO by Raman spectroscopy**. Compared to the graphene monolayer, which is characterized by first order scattering of the $E_{2g}$ mode (G-band), overtone 2D-band, and small defect-induced D-band, GO film has more broadened D- and G-band with similar intensities [30]. A broad D-band corresponds to the high content of oxide groups (hydroxyl, epoxide, etc.) and defects on the lattice basal plane. Reduction mechanisms of GO intend to remove oxide groups in the basal plane and edges of its lattice, which can restore the $sp^2$ domain to some extent and reduce D band intensity. However, using the electrochemical approach, ERGO shows that D-band intensity even increases indicating lattice disorders preservation [31]. Upon GO film deposition on gold, we observed D and G bands of similar intensities ($I_D/I_G$ = 1.05), positioned at 1342 cm$^{-1}$ and 1595 cm$^{-1}$, respectively. Also, ERGO was characterized by D-band at 1343.6 cm$^{-1}$ and G-band at 1602.2 cm$^{-1}$ with preserved $I_D/I_G$ ratio (1.08), see Figure 3A. Certain difference is observed for ERGO on glass surface recorded between IDEs, Figure S3. This means that Raman spectroscopy cannot correlate with electrical characteristics of GO upon reduction, as for monolayer graphene, but it can suggest morphological changes.

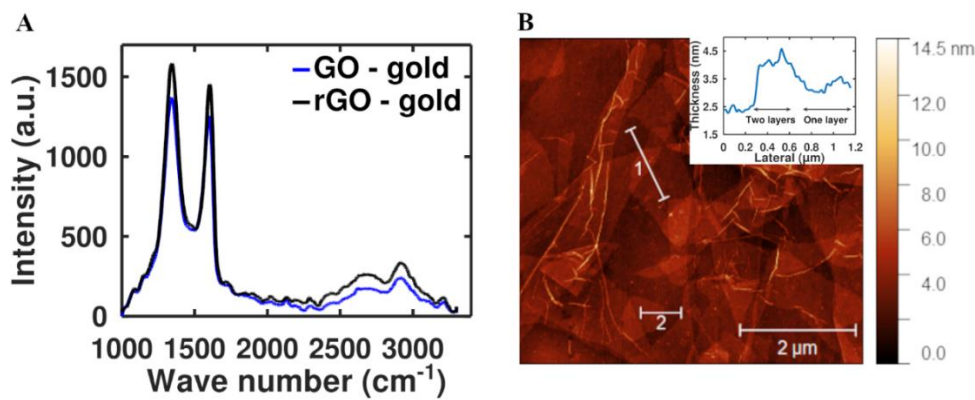

**Figure 3.** (A) Raman spectroscopy of GO/ERGO deposited on gold electrode; (B) AFM micrograph of ERGO film deposited on gold surface with multilayer thickness profile (inset)

**Characterization of GO thin film by AFM**. GO deposition via drop-casting may produce a non-uniform film across the electrode and, therefore, AFM measurements were implemented to study the film morphology and GO flake coverage. GO is deposited from water/NMP mixture without solvent evaporation in the process of electrostatic interaction of positively charged surface and negatively charged GO flakes producing monolayer GO film with islands of multilayer stacks. This is confirmed by AFM micrograph, Figure 3B, presenting GO monolayer flakes deposited randomly on the surface or other GO flakes, as characterized after the reduction. The thickness of a single ERGO flake is estimated as 1.05±0.22 nm. Multilayer cross-section thickness can be used to roughly calculate the number of layers. As depicted in Figure 3B, line 1 shows thickness profile (inset) of the ERGO monolayer stack. From the thickness profile, it was roughly estimated that a stack contains two GO monolayers (thickness of the stack 1.65 nm, thickness of the monolayer 0.84 nm).

*Biosensor development*

Electrochemical reduction removes oxidation groups, which is explained through the reduction peak diminishing in CV voltammogram. However, it does not reconfigure carbon lattice at defect sites and that is confirmed by Raman D peak stability after the reduction, meaning that despite the increased conductivity of the film, the quality of the film remains the same. Nevertheless, functionalization of such a film is more suitable for electrochemical system, compared to as-deposited dielectric GO, because there is much less reactive oxide groups and electrochemical conductivity is more pronounced.

ERGO-IDE surface is functionalized using a standard linker-based protocol [21]. As depicted in Figure 4A, four steps of functionalization are applied and characterized by EIS. Namely, the ERGO surface is firstly functionalized with PBASE, which is followed by addition of the aptamer, which binds covalently through amide linkage to the PBASE. Typical representation of impedance (EIS) measurement is given by the semi-circle Nyquist plot (Figure S4a), where the x-axis is the real part of the total impedance (Re[Z] = Z') and the y-axis is the negative imaginary part of the total impedance (Im[Z] = Z″).

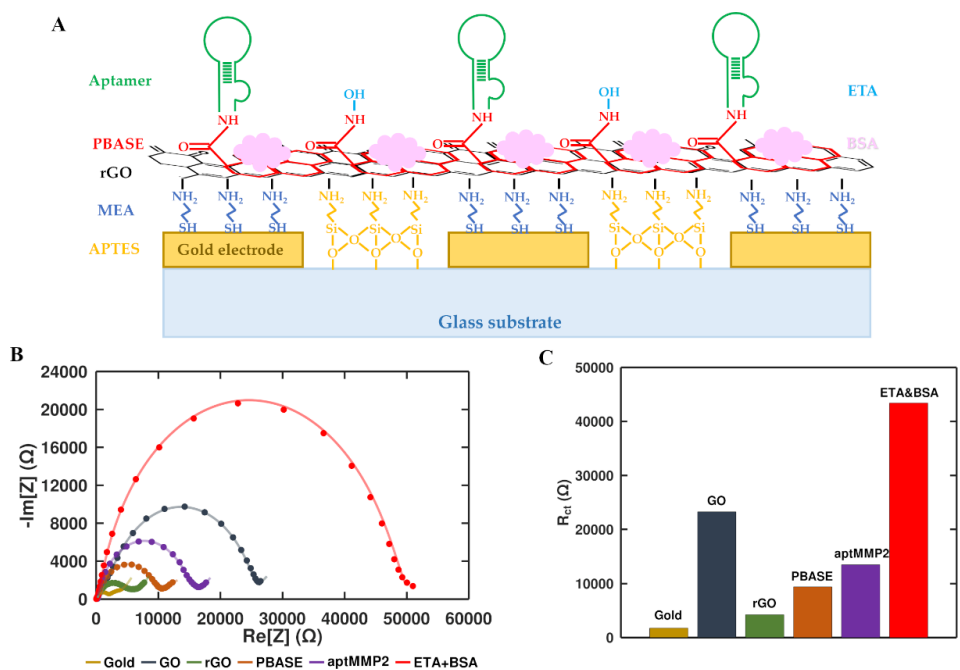

**Figure 4**. (A) Schematic illustration of biosensor functionalization – surface modification by APTES, MEA, and GO followed by PBASE linkers, anti-MMP2 aptamers, and ETA and BSA blocking; (B) Nyquist plots of each biosensor functionalization step indicating overall charge transfer resistance increase, except for the reduction process; (C) Bar plot of $R_{ct}$ values extracted from fitted Nyquist curves from (B)

Introduction of pyrene-based linker on the surface of ERGO-IDE adds a certain barrier for redox charges indicating increase of semi-circle and $R_{ct}$. Pyrene-base linker PBASE interacts with graphene lattice via π-π stacking on one side leaving free ester group on other side for covalent binding of amino-modified aptamers. The addition of specific anti-MMP2 aptamers brings additional resistance at the interface, which is confirmed by $R_{ct}$ increase. Unreacted ester groups were neutralized by ethanolamine via amide bond formation between ester group of PBASE and amino group of ethanolamine. The final step was surface blocking with BSA to prevent unspecific reactions of the surface and the analyte. The Nyquist plot reveals a higher increase in $R_{ct}$ since BSA covers much of the working electrode surface and prevents redox probe to reach the electrode. For clarity, fitted $R_{ct}$ values are presented in Figure 4C from bare gold electrode to fully assembled biosensor. The $R_{ct}$ value of GO film is higher than the almost fully assembled electrode, meaning that reducing GO both increases the conductivity and facilitates the functionalization process.

*Determination of the biosensor performance*

**Sensitivity**. An anti-MMP2 aptamer, with a dissociation constant ($K_d$) of 5.59 nM [20], was selected as a biorecognition element. The block-scheme of sample analysis is presented on Figure S5, and the total duration is about 30 min. In our research, we used an active form of the protein to validate sensor in buffer solution. Impedance spectra was recorded for all range of concentrations from 1 pg mL$^{-1}$ to 100 ng mL$^{-1}$, and a representative set of Nyquist plots is presented in Figure 5A. Hence, the semi-circle increases after each MMP-2 concentration incubation meaning that MMP-2 binds to the aptamer and additionally blocks a path for the redox probe to reach the electrode. Using an equivalent Randles circuit to fit each semi-circle Nyquist plot and determine $R_{ct}$ from the plot (Figure S4b), we were able to define a calibration curve by calculating the signal S, Figure 5B. Hence, the biosensor was able to record a wide dynamic range from 10 pg mL$^{-1}$ to 10 ng mL$^{-1}$. In addition, a limit of detection (LOD) was calculated from the signal-to-noise ratio, where noise is calculated from blank measurement $R_{ct}$ relative error, determining 3.32 pg mL$^{-1}$ as a limit of detection.

The comparison of other biosensing solutions with our approach is given in Table 1, showing that our biosensor has competitive performance and highly improved LOD compared to other aptasensors for MMP-2 detection. Enhanced sensitivity of the constructed biosensor for MMP-2 can be ascribed to the employed methodology for biosensor assembly in comparison to the published procedures, e.g., in the case of the reported monolayer graphene methodology that used maleimide-based click chemistry for aptamer functionalization [12]. Namely, it is known that the enhanced roughness of GO film deposited on IDE, compared to monolayer graphene, can improve the sensitivity [32]. In addition, we used electrochemical reduction optimization to investigate the suitable CV parameters toward achieving high GO film conductivity, which is confirmed by the increase of CV oxidation peak to the level of bare gold electrode or higher (Figure 2A) and reduced system impedance (Figure 2B). Further, ERGO is shown to poses higher heterogeneous electron transfer (HET) rate compared to GO, which is especially observed for ferro/ferricyanide redox probes [33], therefore, improving electrochemical activity of the system. Addition rationale for the enhanced sensitivity can be found in the use of the optimized PBASE-chemistry-coupled aptamers as a biorecognition element in combination with GO-modified IDEs. Recently, it was discussed that aptamer coverage/density at the surface of the nanomaterial can influence its equilibrium dissociation constant [34], which can improve sensitivity of the developed biosensor. On the other hand, maleimide adducts, used in [12], can be degraded and undergo thiol exchange reactions in physiological conditions, which may yield to lower aptamer coverage than in the case of PBASE-chemistry-based functionalized graphene in our work [35]. The main effect on sensitivity is caused by steric blocking and electrostatic repulsion between the negatively charged phosphate groups of the DNA and the redox probe [36]. Moreover, the presence of $NH_2$ groups under the surface of ERGO can increase the electrostatic interaction with the aptamer and protein. Due to complex interactions between aptamer and enzyme that include van der Waals forces, hydrogen bonds, and electrostatic interactions, the multiple synergistic binding of the analyte to the sensor surface and the enhanced binding efficiency in biosensor can be suggested. Hence, we hypothesize that our assay led to improving the overall binding affinity of MMP-2 to the anti-MMP2 aptamers functionalized on ERGO, leading to the increased sensitivity and lowering of LOD.

**Table 1.** Detection of MMP-2 by different electrochemical biosensors

| EC method | Assay & material | Receptor | Linear range | LOD | Ref. |
|---|---|---|---|---|---|
| DPV | PSE/Gr-IDE | Aptamer | 500 - $10^4$ pg mL$^{-1}$ | 6.2 ng mL$^{-1}$ | [12] |
| CA | K-GS@CS@$C_9H_{14}NBF_4$ | Antibody + three ssDNA | $10^{-4}$ –10 ng mL$^{-1}$ | 35 fg mL$^{-1}$ | [37] |
| SWV | ST-gel/CS-AuNPs-Pb/peptides/AuNPs/PAni-gel/GCE | Peptide | 0.001 – 1000 ng mL$^{-1}$ | 0.4 pg mL$^{-1}$ | [38] |
| SWV | peptide2-AgNPs/CB[8]/peptide1-Gold | Peptide | 0.5 pg mL$^{-1}$ – 50 ng mL$^{-1}$ | 0.12 pg mL$^{-1}$ | [39] |
| DPV | RhB/β-CD/PtGaNRs@rGO/GCE | Peptide | 0.1 fg mL$^{-1}$ - 0.1 ng mL$^{-1}$ | 36.57 ag mL$^{-1}$ | [40] |
| EIS | ERGO-IDE | Aptamer | 10 pg mL$^{-1}$ – 10 ng mL$^{-1}$ | 3.32 pg mL$^{-1}$ | This work |

DPV, dynamic pulsed voltammetry; PSE, pyrene-pyridinyl disulphide; CA, chronoamperometry; K-GS, potassium doped graphene sheets; CS, chitosan; SWV, squared-wave voltammetry; ST, sodium tartrate; CS-AuNPs-Pb, Pb-loaded carbon spheres – gold nanoparticles nanocomposite; PAni, polyaniline; AgNPs, silver nanoparticles, CB[8], cucurbit[8]uril; RhB, Rhodamine B; β-CD, beta cyclodextrin; PtGaNRs, platinum-gallium nanorods.

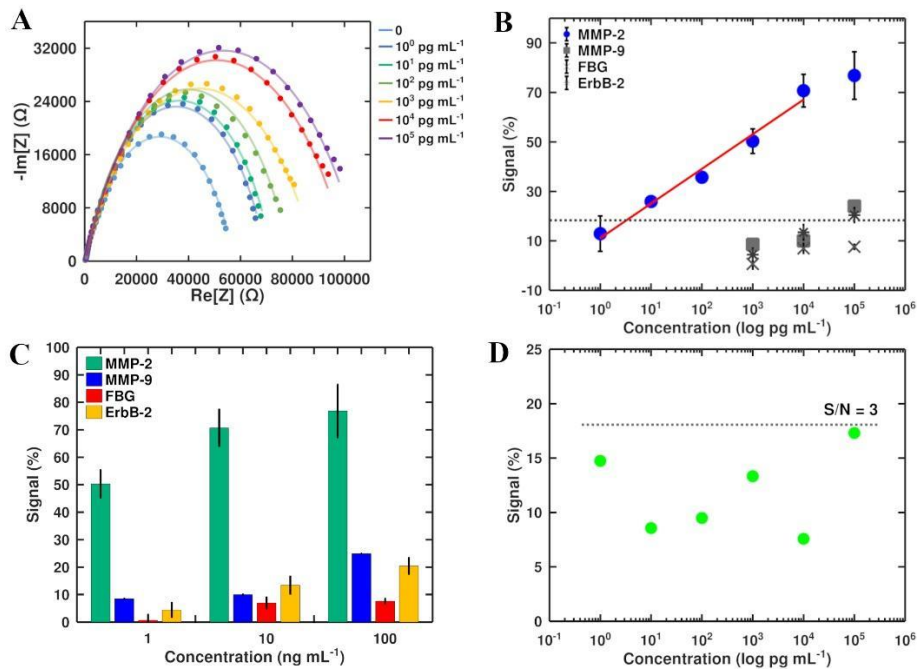

**Figure 5.** (A) Nyquist plots of biosensor response to MMP-2 concentration increase in 1X PBS; (B) Determination of biosensor signal S calibration toward MMP-2 detection (blue) and non-specific target response signals (grey) (dashed grey line is calculated LOD from S/N=3); (C) Biosensor cross-selectivity histogram of response signals for MMP-2, MMP-9, FBG, and ErbB-2; (D) Control biosensor (no aptamers) response toward different MMP-2 concentrations

**Selectivity and regeneration of the biosensor.** The MMP-2 biosensor developed in this research is tested for its selectivity performance in the presence of three proteins: matrix metalloproteinase 9 (MMP-9), fibrinogen (FBG), and tyrosine-protein kinase receptor 2 (ErbB-2) as non-specific analytes. MMP-9 belongs to the same group of gelatinases as MMP-2 and it is also considered as a prognostic cancer biomarker [5].

FBG is recently discussed as a natural inhibitor of MMP-2 [41], while ErbB-2 is a known indicator for breast cancer growth and metastasis [42]. Figure 5C shows that the biosensor can clearly discriminate between specific and non-specific target molecules. Only high concentrations of non-specific target proteins produce a signal comparable to the detection limit of our biosensor (≤20%) (see Fig. 5B). Additionally, we tested the biosensor's specificity of non-aptamer-modified assay, where ERGO-IDEs were modified with PBASE, blocking molecules, but no aptamers. We confirmed that this assay is not able to detect MMP-2 with an expected trend, see Figure 5D. Finally, the aptamer regeneration using NaOH aqueous solution was proposed to regenerate biosensor during the aptamer denaturation in basic solution. The basic environment leads to aptamer unfolding and enzyme release [22,23]. After each set of experiments, electrodes were washed thoroughly with NaOH solution and blank measurements were repeated, giving almost complete signal regeneration revealing the possibility of biosensor reuse, shown in Figure S6.

**Real samples**. MMP-2 detection by the proposed biosensor was tested on the human serum samples as a proof-of-concept. The comparison of the concentrations determined by the ELISA and biosensor showed that our device is able to detect MMP-2 in a complex matrix, such as human serum, see Figure 6. It should be noted that the upper limit of the ELISA kit is 50 ng mL$^{-1}$, and the actual concentration of sample 2 and 3 may be higher than measured.

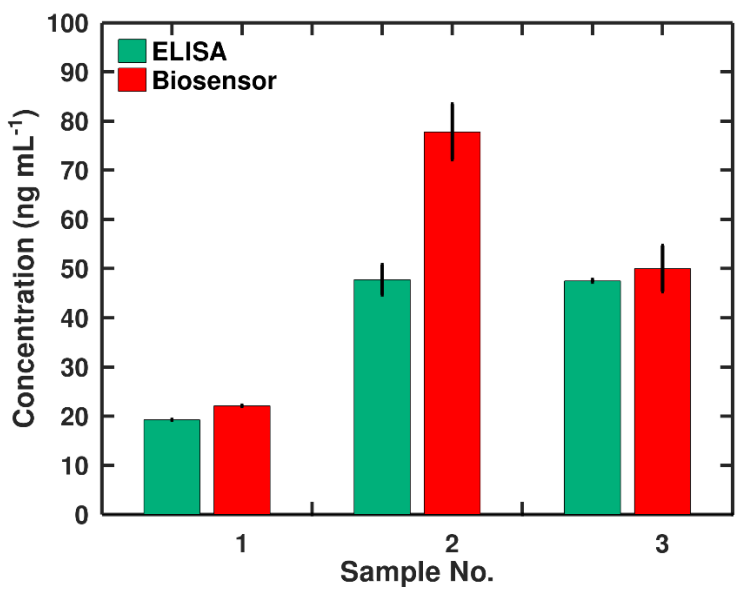

**Figure 6**. Measurements of MMP-2 concentrations in human serum samples obtained from 3 patients and validated by ELISA test (N=2)

**Conclusion**

A novel facile methodology for the development of highly sensitive and selective biosensors for the detection of MMP-2 has been successfully employed and tested. Commercial gold IDEs and its glass substrate are functionalized with positively charged amine groups which enabled efficient electrostatic attachment of GO through drop-casting methodology. Here, a fully electrochemical strategy was further applied to reduce GO thin film to the ERGO conductive films by an optimized CV technique. After employing 20 CV cycles, a highly conductive film was produced, as proven by the increased oxidative peak current in CV or reduced $R_{ct}$ impedance shown by EIS, while the morphology of GO is preserved as confirmed by Raman spectroscopy. Moreover, this system was applied in the development of a biosensor for MMP-2 detection, using pyrene-based modification of ERGO and covalent attachment of specific anti-MMP2 aptamer. With this approach, the high biosensor sensitivity was demonstrated by detecting very low concentration (of active MMP-2 in a standard buffer solution), with the calculated LOD of 3.32 pg mL$^{-1}$. The ERGO-IDE aptasensor further showed good cross-selectivity to MMP-2 over similar enzyme MMP-9 and other proteins, such as fibrinogen and tyrosine-protein kinase receptor 2. The potential of reuse by aptamer regeneration is also demonstrated, which further supports the advantages of the employed methodology. Finally, our biosensor was able to compete with validated standard methods, such as ELISA, in detection of MMP-2 in human serum. The employed technology and novel biosensor exhibit promising characteristics for use in point-of-care devices for MMP-2 detection and applications in the prognosis and early diagnosis of cancer, but also for other pathological processes where MMP-2 overexpression is involved.

**Credit author statement**

**Stefan Jarić**: Conceptualization, Methodology, Investigation, Validation, Formal analysis, Visualization, Writing – original draft; **Silvia Schobesberger**: Methodology, Validation, Formal analysis, Writing – review & editing; **Lazar Velicki**: Resources; **Aleksandra Milovančev**: Resources; **Stanislava Nikolić**: Investigation, Validation, Formal analysis; **Peter Ertl**: Writing - review & editing, Resources, Project administration; **Ivan Bobrinetskiy**: Conceptualization, Methodology, Validation, Writing – review & editing, Supervision; **Nikola Knežević**: Conceptualization, Writing – review & editing, Project administration; Funding Acquisition.

**Declaration of competing interest**

The authors declare that they have no known competing financial interests or personal relationships that could have appeared to influence the work reported in this paper.

**Research ethics**

The authors declare that any aspect of the work covered in this manuscript that has involved human patients has been conducted within the principles of Good Clinical Practice and following the Declaration of Helsinki. The study protocol was approved by the Institutional Review Board (2594-1/3).

**Data availability**

Data will be made available on request.

**Acknowledgements**

This research is funded by Europe Commission's Horizon 2020 Twinning program NANOFACTS [grant No. 952259] (https://doi.org/10.3030/952259). IB acknowledges financial support of Europe Commission's Horizon 2020 Teaming program ANTARES [grant No. 664387] (https://doi.org/10.3030/739570). SJ acknowledges the financial support of Science Fund of the Republic of Serbia IDEAS program MicroLabAptaSens [grant No. 7750276] and the Ministry of Science, Technological Development, and Innovations of the Republic of Serbia [grant No. 451-03-66/2024-03/200358]. LV and AM acknowledge the financial support of the Autonomous Province of Vojvodina—Projects of importance for the development of scientific research activities (2021–2024), [grant No. 142-451-2568/2021-01].